\documentclass{amsart}

\usepackage{amssymb}
\usepackage{enumerate, xspace}
\usepackage{dsfont}
\usepackage{afterpage}
\usepackage{changebar}
\usepackage[dvips]{graphicx}
\usepackage{amsthm}
\usepackage{rotating}

\newif\ifjoel
\ifjoel\fi

\numberwithin{equation}{section}
\newtheorem{thm}{Theorem}[section]

\newenvironment{proofof}[1]{\medskip\noindent
   \textit{Proof of #1:} }{\hfill \qed\par\medskip}

\theoremstyle{definition}
\newtheorem{rem}[thm]{Remark}

\newcommand{\be}{\begin{equation}}
\newcommand{\ee}{\end{equation}}
\newcommand{\bea}{\begin{eqnarray}}
\newcommand{\eea}{\end{eqnarray}}
\newcommand{\bs}{\begin{split}}
\newcommand{\es}{\end{split}}

\newcommand{\ds}{\displaystyle}

\newcommand{\floor}[1]{\lfloor#1\rfloor}

\newcommand{\talpha}{\tilde\alpha}
\newcommand{\tbeta}{\tilde\beta}
\newcommand{\Vn}{|V\rangle}
\newcommand{\Wn}{\langle W|}

\newcommand{\V}[1]{|V_{#1}\rangle}
\newcommand{\W}[1]{\langle W_{#1}|}

\begin{document}

%% \title[Open Two-Species Exclusion Processes]
%% {On Some Classes of Open Two-Species Exclusion Processes}
%% \author{Arvind Ayyer}
%% \address{Arvind Ayyer\\
%% Institut de Physique Th´eorique  IPhT\\
%% CEA Saclay and URA 2306, CNRS\\
%% 91191 Gif-sur-Yvette Cedex, FRANCE.}
%% \email{arvind.ayyer@cea.fr}
%% \author{Joel L. Lebowitz}
%% \address{Joel L. Lebowitz\\
%% Department of Mathematics \\
%% Rutgers University \\
%% 110 Frelinghuysen Road \\
%% Piscataway, NJ 08854, USA\\
%% also Department of Physics \\
%% Rutgers University \\
%% 136 Frelinghuysen Road \\
%% Piscataway, NJ 08854, USA}
%% \email{lebowitz@math.rutgers.edu}
%% \author{Eugene R. Speer}
%% \address{E. R. Speer\\
%% Department of Mathematics \\
%% Rutgers University \\
%% 110 Frelinghuysen Road \\
%% Piscataway, NJ 08854, USA}
%% \email{speer@math.rutgers.edu}

\title[Open Two-Species Exclusion Processes]
  {On Some Classes of Open Two-Species Exclusion Processes} 
\author[Arvind Ayyer]{Arvind Ayyer$^1$} \email{arvind.ayyer@cea.fr}
\author[Joel L. Lebowitz]{Joel L. Lebowitz$^{2,3}$}
\email{lebowitz@math.rutgers.edu} 
\author[Eugene R. Speer]{Eugene R. Speer$^3$} 
\email{speer@math.rutgers.edu}
\keywords{Exclusion processes, 2 species TASEP, open system, 
matrix ansatz, coupling}
\subjclass[2010]{82C22,60K35}
\thanks{\hspace{-\parindent}
1.\hspace{2.9pt}Institut de Physique Th´eorique, IPhT, CEA Saclay,
and URA 2306, CNRS, 91191 Gif-sur-Yvette Cedex, FRANCE\\ 
2.\hspace{2.9pt}Department of Physics, Rutgers University,
136 Frelinghuysen Rd, Piscataway, NJ 08854.\\ 
3.\hspace{2.9pt}Department of Mathematics, Rutgers University, 110
Frelinghuysen Road, Piscataway, NJ 08854.}

\begin{abstract} We investigate some properties of the nonequilibrium
stationary state (NESS) of a one dimensional open system consisting of
first and second class (type~1 and type~2) particles.  The dynamics are
totally asymmetric but the rates for the different permitted exchanges
($1\,0\to0\,1$, $1\,2\to2\,1$, and $2\,0\to0\,2$) need not be equal.  The
entrance and exit rates of the different species can also be different.  We
show that for certain classes of rates one can compute the currents and
phase diagram, or at least obtain some monotonicity properties.  For other
classes one can obtain a matrix representation of the NESS; this
generalizes previous work in which second class particles can neither enter
nor leave the system.  We analyze a simple example of this type and
establish the existence of a randomly located shock  at which the
typical density profiles of all three species are discontinuous.\end{abstract}

\maketitle

\section{Introduction} \label{sec:intro}

The properties of open systems kept in a nonequilibrium stationary state
(NESS) through contact with very large (formally infinite) reservoirs at
different temperatures and chemical potentials is a central issue in
statistical mechanics.  Of particular interest in these systems are the
fluxes of locally conserved quantities, such as particle numbers,
transported by the system from one reservoir to another.  The only
nontrivial systems for which the NESS is known (more or less) explicitly
are one dimensional lattice systems with stochastic dynamics corresponding
to simple exclusion processes or zero range processes \cite{zrp,bdgjl}.  The
latter case is rather exceptional in that the NESS is described fully by a
product measure and thus does not exhibit the long range correlations
expected to be generic for NESS of realistic systems \cite{HS1,Sch}.

In this note we investigate the particle fluxes and densities in the
NESS of the open two-species totally asymmetric simple exclusion process
(2-TASEP).  This is a system on a finite one-dimensional lattice of $L$
sites, in which each site is either empty (empty sites are also referred to
as {\it holes}) or is occupied by a {\it type~1} or a {\it
type~2} particle.  The internal (bulk) dynamics are those in
which a type~1 particle can exchange with a hole or type~2
particle to its right and a type~2 particle with a hole to its right
or a type~1 particle to its left; that is, a type~1 particle at site
$j$ can jump to site $j+1$ unless that site is occupied by another type~1
particle, and a hole at site $j$ can jump to site $j-1$ if that site is
nonempty.  For the moment we allow different rates for each type of
exchange, setting the time scale by normalizing the particle-hole exchange
rate to be 1, but in many of the special cases to be considered below we
will specialize to have equal rates for all bulk exchanges.  We will follow
the convention of \cite{be} and refer to type~1 and type~2 particles as
{\it first class} and {\it second class}, respectively, when the bulk
exchange rates are all equal, as in the original definition of second class
particles \cite{abl}.  Particles
enter the system from the left, i.e., at site~1; type~1 particles can
do so either by filling a hole or by displacing a type~2 particle,
and type~2 particles by filling a hole.  Similarly, particles exit at
the right with the replacement of type~1 particles by either holes or
type~2 particles, and of type~2 particles by holes.

The one-dimensional 2-TASEP dynamics have been studied earlier, primarily
using the so-called {\it matrix ansatz}; see \cite{be} for an extensive
review of this method and its applications.  The case in which all exchange
rates are equal was considered both for the closed system on a ring
\cite{djls} and for the system on the infinite lattice \cite{speer,ffk}.
The case of unequal rates for the system on the ring has also been
discussed \cite{derr,mal}, with a concentration on the situation in which
there is only one type 2 or ``defect'' particle.  The equal-rates dynamics
were also studied, and the NESS obtained, for a semipermeable open system,
i.e., one in which second class particles can neither enter nor leave
\cite{arita,arita0,als}; this restriction of course forces the current of
second class particles to be zero.  In \cite{als} it was shown that the
projection of the NESS on the second class parties (whose total number was
fixed) was given by a Gibbs measure in which the interaction was long range
(logarithmic) but restricted to pairs of neighboring second class particles
(that is, pairs separated only by holes and first class particles).

In the present work the restriction to semipermeable systems is dropped.
This permits us to investigate the dependence of the type~2 particle
current on the different bulk and boundary rates.  Unfortunately, we obtain
explicit results only under various restrictions on the rates.  In
Section~\ref{sec:color} we discuss results obtained from projecting the
two-species TASEP onto a one-species TASEP in two different ways, a process
we call {\it coloring}.  In Section~\ref{sec:couple} we use coupling
arguments to obtain some information on how the currents of the different
species depend on the boundary rates.  In Section~\ref{sec:matrix} we
discuss cases in which the system is solvable by a matrix ansatz,  and
in Section~\ref{sec:speccase}  we describe one such system in which, using
the matrix ansatz, we establish the occurence of a  shock at which  all three
density profiles are discontinuous.
  
Table~\ref{table:rates1} gives our general notation for the rates of the
bulk and boundary processes; in some sections below we will adopt simpler
notation more suited to the restricted rates considered there.  We will let
$J_0$, $J_1$, and $J_2$ denote the signed currents of the various particle
species in the NESS; we always have of course $J_0+J_1+J_2=0$.  Note that
always $J_1\ge0$ and $J_0\le0$, while $J_2$ can have either sign. 
Finally, we remark that, given a specification of the rates, an equivalent
system is obtained by interchanging type~1 particles with holes, left with
right, $u$ with $w$, and $\alpha_i$ with $\beta_i$, $i=1,\ldots,3$.

\begin{table}[hbtp!] 
\begin{tabular}{c| l| c} 
Left (Site 1) & \multicolumn1{c|}{Bulk}  & Right (Site $L$) \\ 
\hline 
$0 \to 1$ with rate $\alpha_1$ & $1\,0 \to 0\,1$ with rate $1$ &
    $1 \to 0$ with rate $\beta_1$ \\ 
$0 \to 2$ with rate $\alpha_2$ & $2\,0 \to 0\,2$ with rate $w$ &
     $1 \to 2$ with rate $\beta_2$\\ 
$2 \to 1$ with rate $\alpha_3$ & $1\,2 \to 2\,1$ with rate $v$ &
     $2 \to 0$ with rate $\beta_3$ 
\end{tabular} 
\medskip
\caption{Rates for the general open two-species process \label{table:rates1}}
\end{table}

\section{Coloring} \label{sec:color} 

In this section we assume that the rates for particle exchanges in the bulk
are independent of the types of particle involved, that is, we take $w=v=1$
in the notation of Table~\ref{table:rates1}.  It is well known that for a
system with these bulk rates and no boundaries, that is, a system on a ring
or on the (doubly) infinite lattice, one may project the system onto a
one-species TASEP system, say consisting of black and white particles, in
two distinct ways: we always color the first class particles black and the
holes white, but may color the second class particles either (all) black or
(all) white; we will refer to the systems arising from these distinct
colorings as the $(1,2+0)$ system and the $(1+2,0)$ system, respectively.
Such coloring arguments have been used in the past to understand the
two-species TASEP on the ring \cite{djls} as well as an open semipermeable
system (in which the second class particles are confined) \cite{als}; in
the latter case, however, they provided only an intuitive guide, since the
boundary rates for the semipermeable system do not satisfy the requirements
for consistency with the coloring discussed in the next paragraph.

To extend fully these colorings to the open system we must assume
appropriate boundary rates.  For the $(1,2+0)$ system we require that first
class particles enter the system at a rate which does not depend on whether
site 1 is occupied by a hole or a second class particle, that is, we must
take $\alpha_1=\alpha_3$.  Similarly for the $(1+2,0)$ system we must take
$\beta_1=\beta_3$.  In this note we consider primarily the case in which
both colorings are possible (but see Remark~\ref{onecolor}), and for
notational simplicity we will then modify the notation of
Table~\ref{table:rates1} and write the transition rates in the following
form:

\begin{table}[hbp!] 
\begin{tabular}{c| c| c} 
Left & Bulk & Right \\ 
\hline 
$0 \to 1$ with rate $\alpha$ & $10 \to 01$ with rate $1$ & $1 \to 0$ with 
rate $\beta$ \\ 
$0 \to 2$ with rate $\gamma$ & $20 \to 02$ with rate $1$ & $2 \to 0$ with 
rate $\beta$\\ 
$2 \to 1$ with rate $\alpha$ & $12 \to 21$ with rate $1$ & $1 \to 2$ with 
rate $\delta$ 
\end{tabular} 
\medskip
\caption{Rates for the  case in which both colorings are possible
     \label{table:rates2}}
\end{table}

When the rates are as in Table~\ref{table:rates2} we can apply to each of
the two colored systems the known results for the open single species TASEP
\cite{dehp}.  For that system with entrance rate $\talpha$ and exit rate
$\tbeta$ the asymptotic ($L\to\infty$) current $\tilde J$ of first class
particles is given by
 \be \label{currtasep}
\tilde J(\talpha,\tbeta)
 = \begin{cases}
\frac 14, 
   & \hbox{if $\talpha,\tbeta \geq \frac 12$ (maximal current region),}\\
\talpha(1-\talpha), 
    & \hbox{if $\talpha < \frac 12$ and $\talpha  \leq \tbeta$,
        (low density region),} \\
\tbeta(1-\tbeta), 
   & \hbox{if $\tbeta < \frac 12$ and  $\tbeta <\talpha$
          (high density region).}
\end{cases}
 \ee
 Similarly, the asymptotic density of particles in the bulk is given by
 \be \label{denstasep}
\tilde\rho(\talpha,\tbeta) = \begin{cases}
\frac 12 & \hbox{if $\talpha,\tbeta \geq \frac 12$ (maximal current),}\\
\talpha &  \hbox{if $\talpha < \frac 12$ and $\talpha  \leq \tbeta$
    (low density),}  \\
\text{linear}(\talpha,1-\talpha) & \hbox{if $\talpha=\tbeta < \frac 12$
  (shock line),}\\
1-\tbeta & \hbox{if $\tbeta < \frac 12$ and  $\tbeta <\talpha$
   (high density),} 
\end{cases}
 \ee
where ``linear$(\alpha,1-\alpha)$'' means the profile is linear with
densities $\alpha$ on the left and $1-\alpha$ on the right.  As is well
known, this linear profile occurs due to the formation of a shock, which
performs an unbiased random walk within the system.  Note that because the
current $\tilde J$ is continuous across the shock line
$\talpha=\tbeta<1/2$ there is no 
need for a separate entry for this region
in \eqref{currtasep}.

 Now if the two-species TASEP has rates given by Table~\ref{table:rates2},
then the $(1,2+0)$ system is a one-species TASEP with entry rate $\alpha$
and exit rate is $\beta+\delta$, so that the current $J_1$ and density
$\rho_1$ in the two-species system are given by
$J_1=\tilde J(\alpha,\beta+\delta)$ and
$\rho_1=\tilde\rho(\alpha,\beta+\delta)$, where $\tilde J$ and $\tilde\rho$
are obtained from \eqref{currtasep} and \eqref{denstasep}.  Similarly, the
$(1+2,0)$ system has entry rate $\alpha+\gamma$ and exit rate $\beta$, so
that in the two-species TASEP we have $J_0=-\tilde J(\alpha+\gamma,\beta)$
and $\rho_0=1-\tilde\rho(\alpha+\gamma,\beta)$.  Finally, $J_2=-J_1-J_0$
and $\rho_2=1-\rho_1-\rho_0$.  Thus the phase diagram of the two-species
system is completely determined by the phase diagrams of the two colored
systems.  We summarize in Table~\ref{table:currpd} the currents and in
Table~\ref{table:denspd} the bulk densities in the two-species system; the
second of these tables has one more row and column than the first for the
reason mentioned above in regard to \eqref{currtasep}--\eqref{denstasep}.
Detailed information about the density profile near the boundary of the
system is obtainable similarly from the results of \cite{dehp}.

\begin{table}[p!]
\vbox to \vsize{\null\vfill
\begin{sideways}
\begin{tabular}{|c||c|c|c|}
\hline 
%$\dfrac{(1,2+0) \rightarrow}{(1+2,0) \downarrow}$ & 
\begin{picture}(3.5,1)\put(0.05,1){\line(2,-1){3.37}}
    \put(2,0.35){$(1,2+0)$} \put(0.3,-0.3){$(1+2,0)$}\end{picture} &
$\substack{\\[6pt]\ds \alpha,\beta+\delta \geq \frac 12 \\ \ds\text{
    (Maximal current)}\\[6pt]}$  & 
$\substack{\\[6pt]\ds \alpha < \frac 12, \alpha  < \beta+\delta \\ \ds \text{
    (Low density)}\\[6pt]}$ &  
$\substack{\\[6pt]\ds \beta+\delta < \frac 12, \beta+\delta  \leq \alpha \\
  \ds \text{ (High density)}\\[6pt]}$ \\ 
\hline \hline
$\substack{\ds \alpha+\gamma,\beta \geq \frac 12 \\ \ds 
   \text{     (Maximal current)}}$ &  
$\substack{\\[6pt] \ds J_1 = \frac 14 \\[6pt] \ds J_0 = -\frac 14,\\[6pt] \ds
  J_2 = 0 \\[6pt]}$ &  
$\substack{ \ds J_1 = \alpha(1-\alpha),\\ \ds J_0 = -\frac 14,\\[6pt]
  \ds J_2 > 0 \\} $&   $X$ \\
\hline
$\substack{\ds \beta < \frac 12, \beta <\alpha+\gamma \\ \ds \text{
    (High density)}}$&  
$\substack{\\[6pt] \ds J_1 = \frac 14, \\[6pt] \ds J_0 = -\beta(1-\beta),
  \\[6pt] \ds J_2 < 0 \\[6pt]}$ &  
$\substack{ \ds J_1 = \alpha(1-\alpha), \\[6pt] \ds J_0 =
  -\beta(1-\beta), \\[6pt] \ds J_2  \text 
{ sign undetermined}
 \\[6pt]}$ & 
$\substack{ \ds J_1 = (\beta+\delta)(1-\beta-\delta), \\[6pt] \ds J_0 =
  -\beta(1-\beta), \\[6pt] \ds J_2 < 0 \\[6pt]}$  
\\
\hline
$\substack{\ds \alpha+\gamma < \frac 12, \alpha+\gamma  \leq \beta \\
  \ds \text{ (Low density)}}$ &  
X & 
$\substack{\\[6pt] \ds J_1 = \alpha(1-\alpha), \\[6pt] \ds J_0 =
  -(\alpha+\gamma)(1-\alpha-\gamma), \\[6pt] \ds J_2 > 0 \\[6pt] }$ &  
$\substack{\\[6pt] \ds J_1 = \alpha(1-\alpha)\ds \\[6pt]
   \ds   J_0 = -\beta(1-\beta)=-J_1\\[6pt] \ds J_2=0
  \\[6pt]}$ \\ 
\hline
\noalign{\vskip20pt}
\multicolumn4{p{6truein}}{\scshape Table~\upshape\ref{table:currpd}. 
     Currents for each class of particle.  Note that the
     order of regions in  the rows and columns is different.
     Always $J_2=-(J_0+J_1)$; the table gives the sign of $J_2$, when this is 
     determined. Regions marked $X$ cannot be realized, and the
     region in the lower right corner 
     can be realized only with $\alpha=\beta$ and $\gamma=\delta=0$.}
\refstepcounter{table}
\label{table:currpd}
\end{tabular}
\end{sideways}
\vfill}
\end{table}

\afterpage{\clearpage}

\begin{table}[p!]
\begin{sideways}
\begin{tabular}{|c||c|c|c|c|}
\hline
%$\dfrac{(1,2+0) \rightarrow}{(1+2,0) \downarrow}$ & 
\begin{picture}(3.5,1)\put(0.05,1){\line(2,-1){3.37}}
    \put(2,0.35){$(1,2+0)$} \put(0.3,-0.3){$(1+2,0)$}\end{picture} &
$\substack{\\[6pt]\ds \alpha,\beta+\delta \geq \frac 12 \\ \ds\text{
    (Maximal current)}\\[6pt]}$ &  
$\substack{\\[6pt]\ds \alpha < \frac 12, \alpha  < \beta+\delta \\ \ds \text{
    (Low density)}\\[6pt]}$ &  
$\substack{\\[6pt]\ds \alpha=\beta+\delta < \frac
  12 \\ \ds \text{ (Shock line)}\\[6pt]}$ &
$\substack{\\[6pt]\ds \beta+\delta < \frac 12, \beta+\delta  < \alpha \\ \ds
  \text{ (High density)}\\[6pt]}$ \\ 
\hline \hline
$\substack{\ds \alpha+\gamma,\beta \geq \frac 12 \\ \ds \text{
    (Maximal current)}}$ &  
$\substack{\\[6pt] \ds \rho_1 = \frac 12,\\[6pt] \ds \rho_0 = \frac 12,\\[6pt]
  \ds \rho_2 = 0 \\[6pt]}$ &  
$\substack{\\[6pt] \ds \rho_1 = \alpha,\\[6pt] \ds \rho_0 = \frac 12,\\[6pt]
  \ds \rho_2 = \frac 12 -\alpha \\[6pt]}$ &  
$X$ & 
$X$ \\
\hline
$\substack{\ds \beta < \frac 12, \beta <\alpha+\gamma \\ \ds \text{
    (High density)}}$ &  
$\substack{\\[6pt] \ds \rho_1 = \frac 12,\\[6pt] \ds \rho_0 = \beta,\\[6pt]
  \ds \rho_2 = \frac 12-\beta \\[6pt]}$ &  
$\substack{ \ds \rho_1 = \alpha,\\[6pt] \ds \rho_0 = \beta,\\[6pt] \ds
  \rho_2 = 1-\alpha-\beta \\[6pt]}$ &  
$\substack{ \ds \rho_1 = \text{linear}(\alpha,1-\alpha),\\[6pt] \ds 
  \rho_0 = \beta,\\[6pt] \ds  \rho_2 =
  \text{linear}(1-\alpha-\beta,\alpha-\beta) \\[6pt]} $ &
$\substack{ \ds \rho_1 = 1-\beta-\delta,\\[6pt] \ds \rho_0 =
  \beta,\\[6pt] \ds \rho_2 = \delta \\[6pt]}$ \\ 
\hline
$\substack{\ds \alpha+\gamma =\beta < \frac 12 \\ \ds \text{ (Shock line)}}$ & 
 $X$ & 
 $\substack{ \ds \rho_1 = \alpha,\\[6pt] \ds \rho_0 =
   \text{linear}(1-\beta,\beta), \\[6pt] \ds  \rho_2 =
   \text{linear}(\beta-\alpha,1-\alpha-\beta) \\[6pt]}$ &  
$\substack{\\[6pt] \ds \rho_1 = \text{linear}(\alpha,1-\alpha),\\[6pt] 
  \ds \rho_0 = \text{linear}(1-\alpha,\alpha),\\[6pt] 
 \ds \rho_2=0 \\[6pt]}$ &
$X$ \\ 
\hline
$\substack{\ds \alpha+\gamma < \frac 12, \alpha+\gamma  < \beta \\ \ds
  \text{ (Low density)}}$ &  
$X$ & 
$\substack{\\[6pt] \ds \rho_1 = \alpha,\\[6pt] \ds \rho_0 =
  1-\alpha-\gamma,\\[6pt] \ds \rho_2 = \gamma \\[6pt]}$ &  
$X$ & 
$X$ \\
\hline
\noalign{\vskip20pt}
\multicolumn5{p{7.8truein}}{\scshape Table~\upshape\ref{table:denspd}. 
     Currents for each class of particle. Note that the
     order of regions in  the rows and columns is different. 
     Regions marked $X$ cannot be realized, and 
     the region in which both subsystems are on their
     respective shock lines can be realized only if $\alpha=\beta$ and 
     $\gamma=\delta=0$.}
\refstepcounter{table}
\label{table:denspd}
\end{tabular} 
\end{sideways}
%\bigskip\caption{Bulk density for each class of particle\label{table:denspd}}
\end{table}

\begin{rem}\label{rem:tables} Examination of Tables~\ref{table:currpd} and
\ref{table:denspd} shows several interesting features of these systems.

 \smallskip\noindent
 (a) In regions of parameter space in which neither colored subsystem is on
its shock line the densities are uniform in the bulk.  In fact we see from
Table~\ref{table:denspd} that, given any $\rho=(\rho_0,\rho_1,\rho_2)$ with
$\sum_i\rho_i=1$, there exist parameters $\alpha,\beta,\gamma,\delta$ which
yield the $\rho_i$ as the (constant) bulk densities.  Recall \cite{speer}
that for any such $\rho$ there exists a unique steady state $\mu^{\rho}$
for the 2-TASEP dynamics on an infinite lattice having these as densities.
It is then easy to see that in the open system with bulk densities $\rho$
the {\it local state} \cite{ligg2,als} at any point a fraction $x$,
$0<x<1$, of the way through the system will be $\mu^{\rho}$; that is, in
the $L\to\infty$ limit the NESS near the site $i=\floor{xL}$ (where
$\floor\xi$ denotes the integer part of $\xi$) will look like $\mu^{\rho}$.

 \smallskip\noindent
 (b) When both colored subsystems are in the maximal current region, i.e., when
$\alpha,\beta\ge1/2$, second class particles are excluded from the bulk and
their current vanishes.  It follows from the results of \cite{dehp},
however, that second class particles are present in the system, with a
density decreasing as the inverse square root of the distance from either 
boundary.

 \smallskip\noindent
 (c) When both colored systems are on their respective shock lines,
necessarily $\alpha=\beta$ and $\gamma=\delta=0$; the system is then a
one-species TASEP, again on the shock line.

 \smallskip\noindent
 (d) When the $(1,2+0)$ system is on its shock line and the $(1+2,0)$
system is not, the latter must be in its high density phase.  In that case
there is a uniform density $\beta$ of holes in the system, together with a
shock, with random location uniformly distributed over the system,
separating a region on the left with densities $\rho_0'=\beta$,
$\rho'_1=\alpha$, and $\rho'_2=1-\alpha-\beta$, from a region on the right
with $\rho_0''=\beta$, $\rho''_1=1-\alpha$, and $\rho''_2=\alpha-\beta$.
The local state a fraction $x$ of the way through the system is then the
superposition $x\mu^{\rho'}+(1-x)\mu^{\rho''}$.  Similar conclusions hold
when the $(1+2,0)$ system is on its shock line and the $(1,2+0)$ system is
in its low density phase.

\end{rem}

\begin{rem}\label{onecolor} One can, of course, also discuss values of the
boundary rates which respect one but not both of the colorings.  Much less
can then be said, but there is one simple case in which the currents can be
completely determined.  Suppose that $\alpha_1=\alpha_3=\alpha$, so that
the $(1,2+0)$ coloring is possible; then since the exit rate of type~1
particles is $\beta_1+\beta_2$ the current $J_1$ is related to the density
$\rho_1(L)$ of type~1 particles at site $L$ by
 \be
J_1=(\beta_1+\beta_2)\rho_1(L).
 \ee
 Now suppose further that $\beta_3=0$, so that type~2 particles
cannot exit at the right boundary and holes can enter there only when a
type~1 particle exits.  The currents across the right boundary (and
thus across any bond) must then satisfy
 \be
 J_0=-\beta_1\rho_1(L)=-\frac{\beta_1}{\beta_1+\beta_2}J_1,\qquad
 J_2=-\beta_2\rho_1(L)=-\frac{\beta_2}{\beta_1+\beta_2}J_1.
 \ee
Note that these results apply for any bulk rate $w$, although we still
require that $v=1$ in order that the $(1,2+0)$ coloring be possible.
\end{rem}

\section{Coupling and monotonicity of currents} \label{sec:couple} 

We do not know how to calculate the currents of the various species in our
model for general rates, but in some cases it is possible to show by means
of coupling arguments that the currents depend monotonically on the rates.
The next two theorems give results of this nature.

\begin{thm}\label{thm:monoj10} Suppose that $v=w=1$.  Then in the region of
parameter space in which $\alpha_3\ge\alpha_1$ and $\beta_3\ge\beta_1$:

 \smallskip\noindent
 (a) The current $J_1$ is monotonically increasing in
$\alpha_1$, $\alpha_2$, and $\alpha_3$; it is monotonically decreasing in
$\beta_3$ and under increase of $\beta_1$ when $\beta_1+\beta_2$ is
constant.

 \smallskip\noindent
 (b) The (negative) current $J_0$ is monotonically increasing in magnitude
as a function of $\beta_1$, $\beta_2$, and $\beta_3$; it is monotonically
decreasing in magnitude in $\alpha_3$ and under increase of $\alpha_1$ when
$\alpha_1+\alpha_2$ is constant.  \end{thm}

\begin{thm}\label{thm:monoj2} Suppose that $v=w=1$.  Then:

 \smallskip\noindent
 (a) If $\alpha_1=\alpha_3$ then the current $J_2$ is monotonically
increasing in $\alpha_2$, in $\beta_3$, and under increase of $\beta_1$
when $\beta_1+\beta_2$ is constant.

 \smallskip\noindent
 (b) If $\beta_1=\beta_3$ then the current $J_2$ is monotonically
decreasing in $\beta_2$, in $\alpha_3$, and under increase of $\alpha_1$
when $\alpha_1+\alpha_2$ is constant.  \end{thm}

To develop the proofs of these theorems, given later in this section, we
introduce the natural order $0\prec2\prec1$ on the three species of
particles, and if $\tau$ and $\tau'$ are two configurations of our system
we write $\tau\preceq\tau'$ if $\tau_i\preceq\tau_i'$ for $i=1,\ldots,L$.
To {\it couple} two evolving systems, denoted as primed and unprimed, is to
define a joint dynamics for the pair $(\tau'(t),\tau(t))$ in such a way
that the induced dynamics on each separate system is that of
Table~\ref{table:rates1}, with rates $v',w',\alpha'_i,\beta'_i$ and
$v,w,\alpha_i,\beta_i$, respectively. We are interested only in order
preserving couplings, for which if $\tau'(0)\succeq\tau(0)$ then
$\tau'(t)\succeq\tau(t)$ for all $t\ge0$.

We now define the coupling in the bulk and on the boundaries.

 \smallskip\noindent
 {\bf Bulk:} For this coupling  it is necessary that all the bulk
rates for the two systems be the same:
 \be
 v=v'=w=w'=1.\label{eq:bulkc}\\
 \ee
 Under this assumption we define the coupled dynamics by supposing that a
Poisson alarm clock which rings with rate 1 is associated with each bond,
and that when the alarm rings any possible exchange across that bond, in
either the primed or unprimed system, takes place.  Equivalently one may
describe the dynamics by giving rates for the coupled process: if
$(\tau_i',\tau_i)\ne(\tau_{i+1}',\tau_{i+1})$ then at rate 1,
 \bea\label{eq:ratesbc}
 (\tau_i',\tau_i)(\tau_{i+1}',\tau_{i+1}) 
   &\to& (\tau_{i+1}',\tau_{i+1}) (\tau_i',\tau_i)\quad
    \hbox{if $\tau_i'\succeq\tau_{i+1}'$ 
   and $\tau_i\succeq\tau_{i+1}$,}\nonumber\\
 (\tau_i',\tau_i)(\tau_{i+1}',\tau_{i+1}) 
   &\to& (\tau_i',\tau_{i+1}) (\tau_{i+1}',\tau_i)\quad
    \hbox{if $\tau_i'\prec\tau_{i+1}'$ 
   and $\tau_i\succeq\tau_{i+1}$,}\\
 (\tau_i',\tau_i)(\tau_{i+1}',\tau_{i+1}) 
   &\to& (\tau_{i+1}',\tau_i) (\tau_i',\tau_{i+1})\quad
    \hbox{if $\tau_i'\succeq\tau_{i+1}'$ 
   and $\tau_i\prec\tau_{i+1}$,}\nonumber
 \eea
 It is easy to see that no coupling of this general sort
 will succeed if \eqref{eq:bulkc} does not hold.  

 \smallskip\noindent
 {\bf Boundaries:} As an example we begin by considering the particular
case in which the state on site $L$ is $(\tau_L',\tau_L)=(1',1)$.  We must
assign rates to the five allowed transitions on site $L$ arising from the
exit of the first class particle in one or both of the coupled systems,
that is, transitions from $(1',1)$ to $(0',0)$, $(2',0)$, $(2',2)$,
$(1',0)$, or $(1',2)$; on the other hand, transitions to $(0',1)$,
$(0',2)$, or $(2',1)$, which would violate the ordering of the two
configurations, must not occur.  The total rate for each transition in each
of the two coupled systems must be given by Table~\ref{table:rates1}, so
that for example the rate for $1'\to 2'$, which is the sum of the rates for
$(1',1)\to(2',0)$ and $(1',1)\to(2',2)$ must be $\beta_2'$, the sum of the
rates for $(1',1)\to(0',0)$, $(1',1)\to(2',0)$ and $(1',1)\to(1',0)$, must
be $\beta_1$, etc.  There are thus four equations to be satisfied by five
rates; it is easy to see that a solution exists with nonnegative rates if and
only if $\beta_1\ge\beta_1'$ and $\beta_1+\beta_2\ge\beta_1'+\beta_2'$, and
that there is then in general one free parameter.

All other boundary transitions may be analyzed similarly; the resulting
rates are given in Table~\ref{table:ratesc}.  In this table $x$, $y$, $s$,
and $t$ are parameters which may be chosen freely subject to the condition
that all the resulting rates be nonnegative, that is, to the inequalities
 \be\label{eq:parac}\begin{split}
 \max\{0,\alpha_2-\alpha_2'\}&\le x\le \min\{\alpha_2,\alpha_1'-\alpha_1\},\\
 0&\le y\le \min\{\alpha_2,\alpha_3'-\alpha_1\},\\
 \max\{0,\beta_2'-\beta_2\}&\le s\le \min\{\beta_2',\beta_1-\beta_1'\},\\
 0&\le t\le \min\{\beta_2',\beta_3-\beta_1'\}.
 \end{split}\ee
 The inequalities \eqref{eq:parac} will have a solution, and all rates of
Table~\ref{table:ratesc} will be nonnegative, if and only if the parameters
of the two systems satisfy
 \begin{gather}
\alpha_1'\ge\alpha_1,\quad\alpha_1'+\alpha_2'\ge\alpha_1+\alpha_2,\quad
\alpha_3'\ge\alpha_3,\quad\alpha_3'\ge\alpha_1,\label{eq:leftc}\\
\beta_1'\le\beta_1,\quad\beta_1'+\beta_2'\le\beta_1+\beta_2,\quad
\beta_3'\le\beta_3,\quad\beta_1'\le\beta_3.\label{eq:rightc} \end{gather}
In summary: an order preserving coupling of the two systems is possible if
\eqref{eq:bulkc}, \eqref{eq:leftc}, and \eqref{eq:rightc} are satisfied,
and is then given by Table~\ref{table:ratesc}.

\begin{table}[htp!] 
\begin{tabular}{cc|c c} 
Left & Rate &  Right & Rate \\ 
\hline 
$(0',0) \to (1',1)$& $\alpha_1$ &
      $(1',1) \to (0',0)$& $\beta_1'$ \\ 
$(0',0) \to (1',2)$& $x$ &
      $(1',1) \to (2',0)$& $s$ \\
$(0',0) \to (1',0)$& $\alpha_1'-\alpha_1-x$ &
       $(1',1) \to (2',2)$& $\beta_2'-s$ \\
$(0',0) \to (2',2)$& $\alpha_2-x$ &
      $(1',1) \to (1',0)$& $\beta_1-\beta_1'-s$ \\
$(0',0) \to (2',0)$& $\alpha_2'-\alpha_2+x$ &
      $(1',1) \to (1',2)$& $\beta_2-\beta_2'+s$ \\
$(2',0) \to (1',1)$& $\alpha_1$ &
      $(1',2) \to (0',0)$& $\beta_1'$\\
$(2',0) \to (1',2)$& $y$ &
      $(1',2) \to (2',0)$& $t$\\
$(2',0) \to (1',0)$& $\alpha_3'-\alpha_1-y$ &
      $(1',2) \to (2',2)$& $\beta_2'-t$\\
$(2',0) \to (2',2)$& $\alpha_2-y$ &
      $(1',2) \to (1',0)$& $\beta_3-\beta_1'-t$\\
$(2',2) \to (1',1)$& $\alpha_3$ &
      $(1',0) \to (0',0)$& $\beta_1'$\\
$(2',2) \to (1',2)$& $\alpha_3'-\alpha_3$ &
      $(1',0) \to (2',0)$& $\beta_2'$\\
$(1',0) \to (1',1)$& $\alpha_1$ &
      $(2',2) \to (0',0)$&$\beta_3'$\\
$(1',0) \to (1',2)$& $\alpha_2$ &
      $(2',2) \to (2',0)$& $\beta_3-\beta_3'$\\
$(1',2) \to (1',1)$& $\alpha_3$ &
      $(2',0) \to (0',0)$& $\beta_3'$
\end{tabular} 
\medskip
\caption{Rates for the coupled  process \label{table:ratesc}}
\end{table}

\begin{rem}\label{rem:symmc}(a) Note in particular that the
marginal rates for boundary transitions in either the primed or unprimed
system are those of Table~\ref{table:rates1} separately in each boundary
configuration.  For example, the rate for $1'\to 2'$ at the right boundary
is $\beta_2'$ if $(\tau_L',\tau_L)=(1',1)$, as discussed in the example
above, and also if $(\tau_L',\tau_L)=(1',2)$ or $(\tau_L',\tau_L)=(1',0)$.

 \smallskip\noindent
 (b) The left-right, particle-hole symmetry described in
Section~\ref{sec:intro} can be extended to the coupled system by also
interchanging the primed and unprimed systems.\end{rem}

From now on we will write $J_i'$ and $J_i$ for the currents in the primed
and unprimed systems.

\begin{thm}\label{thm:Jineq} (a) Suppose that the coupling is possible,
i.e., that \eqref{eq:bulkc}--\eqref{eq:rightc} hold, and that in addition
 \be\label{eq:restrict1}
\beta_1+\beta_2=\beta_1'+\beta_2'.
 \ee
Then the currents of first-class particles satisfy $J_1'\ge J_1$.

 \smallskip\noindent
 (b) Similarly, $J_0'\le J_0$ when $\alpha_1+\alpha_2=\alpha_1'+\alpha_2$.
\end{thm}

\begin{proof} (a) By \eqref{eq:parac} and \eqref{eq:restrict1} we must have
$s=\beta_1-\beta_1'=\beta_2'-\beta_2$ in the coupling, so that the rates
are zero for the processes $(1',1) \to (1',0)$ and $(1',1) \to (1',2)$ at
the right end of the system.  Thus a type $1$ particle cannot exit the
system unless a type $1'$ does also, and this implies the result.  For (b)
one may give a similar argument or appeal to the symmetry of
Remark~\ref{rem:symmc}(b). \end{proof}

We can now prove the first  monotonicity result given above.

 \begin{proofof}{Theorem~\ref{thm:monoj10}} We show monotonicity of $J_1$
in $\alpha_1$; all other cases are similar.  If
$\alpha_1,\alpha_2,\alpha_3$ and $\beta_1,\beta_2,\beta_3$ satisfy
$\alpha_3\ge\alpha_1$ and $\beta_3\ge\beta_1$, and $\alpha_1'$ satisfies
$\alpha_3\ge\alpha_1'\ge\alpha_1$, then by defining $\alpha_2'=\alpha_2$,
$\alpha_3'=\alpha_3$, and $\beta_i'=\beta_i$, $i=1,2,3$, we obtain two
systems to which Theorem~\ref{thm:Jineq} may be applied.  \end{proofof}

\begin{rem} We have no evidence that the restrictions in
Theorem~\ref{thm:monoj10} that $\alpha_3\ge\alpha_1$ and
$\beta_3\ge\beta_1$ in are in fact necessary for the monotonicity; they may
well be an artifact of the method of proof.  \end{rem}

We now turn to the proof of Theorem~\ref{thm:monoj2}, for which we
introduce a more restrictive coupling with a larger set of excluded pairs.
Specifically, the coupling has the property that if the initial
configuration of the coupled systems contains none of the pairs
$(\tau_i',\tau_i)=(0',2)$, $(0',1)$, $(2',1)$, $(1',0)$ or $(1',2)$ then
these pairs are also absent from configurations $(\tau'(t),\tau(t))$ for all
$t\ge0$.  For this coupling the parameters must satisfy 
 \begin{gather}
 u=u'=w=w'=1,\label{eq:bulkc2}\\
\alpha_1'=\alpha_1=\alpha_3=\alpha_3',\quad\alpha_2'\ge\alpha_2,
   \label{eq:leftc2}\\
\beta_1'\le\beta_1,\quad\beta_1'+\beta_2'=\beta_1+\beta_2,\quad
   \beta_3'\le\beta_3.\label{eq:rightc2}
  \end{gather}
 Rates for the coupled process are then given in Table~\ref{table:ratesc2}.

 Note that in \eqref{eq:bulkc2}--\eqref{eq:rightc2} we have added
conditions to those required for the previous coupling (that is, to
\eqref{eq:bulkc}, \eqref{eq:leftc}, and \eqref{eq:rightc}) but also omitted
the conditions $\alpha_3'\ge\alpha_1$ and $\beta_1'\le\beta_3$.  Note also
that \eqref{eq:leftc2} implies that both the primed and unprimed systems
can be given the $(1,2+0)$ coloring discussed in Section~\ref{sec:color}
and hence that $J_1'=J_1$; this also follows from the fact that particles
1' and 1 appear in the coupled system only in the pair $(1',1)$.  Thus any
inequality between $J_2'$ and $J_2$ implies the opposite inequality between
$J_0'$ and $J_0$.

\begin{table}[htp!] 
\begin{tabular}{cc|c c} 
Left & Rate &  Right & Rate \\ 
\hline 
$(0',0) \to (1',1)$& $\alpha_1$ &
      $(1',1) \to (0',0)$& $\beta_1'$ \\ 
$(0',0) \to (2',2)$& $\alpha_2$ &
      $(1',1) \to (2',0)$& $\beta_2'-\beta_2$ \\
$(0',0) \to (2',0)$& $\alpha_2'-\alpha_2$ &
      $(1',1) \to (2',2)$& $\beta_2$ \\
$(2',0) \to (1',1)$& $\alpha_1$ &
      $(2',2) \to (0',0)$&$\beta_3'$\\
$(2',0) \to (2',2)$& $\alpha_2$ &
      $(2',2) \to (2',0)$& $\beta_3-\beta_3'$\\
$(2',2) \to (1',1)$& $\alpha_3$ &
      $(2',0) \to (0',0)$& $\beta_3'$
\end{tabular} 
\medskip
\caption{Rates for the coupled  process giving monotonicity of 
  $J_2$\label{table:ratesc2}}
\end{table}

\begin{thm}\label{thm:J2ineq} Suppose that
  \eqref{eq:bulkc2}--\eqref{eq:rightc2} hold.   Then:

 \smallskip\noindent
 (a) If $\beta_i=\beta'_i$ for $i=1,2,3$ then $J_2'\ge J_2$.

 \smallskip\noindent
 (b) If $\alpha_2=\alpha_2'$ then $J_2'\le J_2$.
\end{thm}

\begin{proof} Under the hypothesis of (a) no $(2',0)$ pair can enter the
system at the right boundary, while under (b) no $(2',0)$ pair can enter
the system at the left boundary.  \end{proof}

\begin{proofof}{Theorem~\ref{thm:monoj2}} (a) is obtained form
Theorem~\ref{thm:J2ineq} as in the proof of Theorem~\ref{thm:monoj10}, and
then (b) follows from the symmetry of Remark~\ref{rem:symmc}(b).
\end{proofof}

One can check easily that these results are confirmed by those of
Table~\ref{table:currpd}, when both colorings are possible; in that case
they imply that $J_2$ should be increasing in $\beta$ and $\gamma$ and
decreasing in $\alpha$ and $\delta$.

\section{Matrix solutions} \label{sec:matrix}

In this section we observe that for various special choices of the boundary
rates of Table~\ref{table:rates1} the probabilities of configurations of
the open two-species TASEP can be computed using a variant of the usual
matrix product ansatz. Specifically, this means that for these rates 
there exist matrices
$X_0$, $X_1$, and $X_2$ and vectors $\Vn$ and $\Wn$ such that the
probability of the configuration $\tau_1\ldots\tau_L$, where $\tau_i$ is 0,
1, or 2, is given by
 \be\label{eq:ss}
 \mu_L(\tau_1,\ldots,\tau_L)=Z_L^{-1}w_L(\tau_1,\ldots,\tau_L);
 \ee
here the weight $w_L$ is given by 
 \be\label{eq:weight}
 w_L(\tau_1,\ldots,\tau_L)= \Wn X_{\tau_1}X_{\tau_2}\cdots X_{\tau_L}\Vn,
 \ee
 and the normalizing factor $Z_L$ by
 \be\label{eq:Z}
 Z_L=\sum_{\tau_1,\ldots,\tau_L}w_L(\tau_1,\tau_2,\ldots,\tau_L)
    = \Wn(X_0+X_1+X_2)^L\Vn.
 \ee
 Such matrix solutions exist for two different families of boundary rates,
which we call {\it permeable} and {\it semipermeable}.  We can impose
either type of boundary condition at either end of the system, leading to
four distinct families of open systems.    

We now describe the rates (summarized in
Table~\ref{table:rates3}) and the necessary algebraic properties of the
matrices $X_i$ and vectors $\Vn,\Wn$ for these systems.

\smallskip\noindent
{\bf Bulk:} The exchange rates in the bulk are those of
Table~\ref{table:rates1}, although we shall on occasion specialize to take
$v$ and/or $w$ equal to 1.  For the weights \eqref{eq:ss} to be stationary the
matrices $X_0$, $X_1$, $X_2$ should satisfy
 \be \label{algB}
X_1X_0=X_1+X_0,\qquad X_1X_2=\frac1v X_2,\qquad X_2X_0=\frac1w X_2,
 \ee
 the usual algebraic rules for the 2-TASEP \cite{derr,mal}.

 \smallskip\noindent
 {\bf Semipermeable boundaries:} A semipermeable boundary is one through
which type~2 particles cannot pass, that is, a
semipermeable left boundary has $\alpha_2=\alpha_3=0$ and a semipermeable
right boundary $\beta_2=\beta_3=0$.  A semipermeable boundary is thus
characterized by a single rate, which we will write as $\alpha=\alpha_1$
for a left boundary and $\beta=\beta_1$ for a right boundary.  The vectors 
$\Wn$ for the left boundary and $\Vn$ for the right
boundary should satisfy 
 \begin{eqnarray}
 \hbox{Semipermeable left boundary:}&& \Wn X_0
    =\ds\frac1\alpha\Wn,\label{algSPl}\\
 \hbox{Semipermeable right boundary:}&& X_1\Vn
     =\ds\frac1\beta\Vn.\label{algSPr}
 \end{eqnarray}
 The open system with two semipermeable boundaries and with $w=v=1$ was
studied in \cite{arita0}, \cite{arita}, and \cite{als}.

 \smallskip\noindent
 {\bf Permeable boundaries:} At a permeable boundary type~2 particles
can enter and leave the system, but the rates  are constrained by
 \begin{eqnarray}
 \hbox{Permeable left boundary:}&&\alpha_1+\alpha_2=w, \label{consPl}\\
 \hbox{Permeable right boundary:}&&\beta_1+\beta_2=v. \label{consPr}
 \end{eqnarray}
 We will choose a convenient parametrization of such rates which will be
discussed further below.  At the left boundary we write $\alpha_3=a$ for
the rate of exit of type~2 particles; there is a second free parameter,
$z$, which plays the role of a fugacity for type~2 particles, and
$\alpha_2=waz$, $\alpha_1=w(1-az)$.  At the right boundary we write
similarly $\beta_3=b$, $\beta_2=vbz$, and $\beta_1=v(1-\beta z)$.  If both
boundaries are permeable then the parameter $z$ must be the same on the
left and the right, and to insure nonnegative rates these parameters must
satisfy
 \be
az \le 1,\qquad bz\le 1.
 \ee
 The vectors
vectors $\Vn$ and $\Wn$ used to construct the steady state should satisfy
 \begin{align}
 \hbox{Permeable left boundary:}&& 
   \Wn X_0&=\ds\frac1w\Wn,&\Wn X_2&=z\Wn, \label{algPl}\\
 \hbox{Permeable right boundary:}&&
    X_1\Vn&=\ds\frac1v\Vn,&  X_2\Vn&=z\Vn. \label{algPr}
 \end{align}

\begin{table}[htp!] 
\begin{tabular}{ccc|cc|ccc} 
\multicolumn3{c|}{Left}  & \multicolumn2{c|}{Bulk}  
   & \multicolumn3{c}{Right}\\
&SP&P&&
   &&SP&P\\
\hline
$0 \to 1$& $\alpha$ & $w(1-az)$ &
    $1\,0 \to 0\,1$& $1$ &
    $1 \to 0$& $\beta$&  $v(1-bz)$ \\ 
$0 \to 2$& 0 & $waz$ &
    $2\,0 \to 0\,2$& $w$ &
     $1 \to 2$& 0&  $vbz$\\ 
$2 \to 1$& 0 & $a$ &
 $1\,2 \to 2\,1$& $v$ &
     $2 \to 0$& $0$& b\\[6pt]
\hline
\noalign{\vskip-12pt}&&&&&&&\\
[6pt]$\Wn$&$\W{\alpha}$&$\W{w}$&&&$\Vn$&$\V{\beta}$&$\V{v}$
\end{tabular} 
\bigskip
\caption{Rates for  permeable (P) and
  semipermeable (SP) boundary conditions.  The last line shows the boundary
  vectors $\Wn$ and $\Vn$ to be used in \eqref{eq:ss}. \label{table:rates3}}
\end{table}

It is a straightforward exercise, using standard arguments, to verify that
if there exist matrices and vectors $X_0$, $X_1$, $X_2$, $\Vn$, and $\Wn$
satisfying \eqref{algB}, either \eqref{algSPl} or \eqref{algPl}, and either
\eqref{algSPr} or \eqref{algPr}, then \eqref{eq:ss} defines an invariant
measure for the corresponding dynamics.  We establish this existence below.
One striking consequence is that if the system has one or more permeable
boundaries then the parameters $a$ and/or $b$, which control the entry and
exit of type~2 particles, do not appear in the algebraic conditions
\eqref{algB} and \eqref{algPl} and/or \eqref{algPr}, so that the steady
state of the system is independent of these parameters.

To establish the existence, recall \cite{dehp} that there exist matrices
$X_0$ and $X_1$, and vectors $\langle W_\alpha|$ and $|V_\beta\rangle$
satisfying
 \begin{gather}X_1 X_0  =  X_1+X_0, \\
\langle W_\alpha | X_0  =  \frac{1}{\alpha} \langle W_\alpha|, \qquad
X_1 |V_\beta \rangle  =  \frac{1}{\beta} |V_\beta \rangle \\
 \langle W_\alpha|V_\beta \rangle
    ={\frac{\alpha\beta}{\alpha+\beta-1}},\qquad \alpha+\beta>1. \label{eq:ip}
 \end{gather}
 We now define
 \be\label{eq:defx2}
 X_2=z\frac{w+v-1}{wv}\V{v}\W{w};
 \ee
 then the matrices $X_0$, $X_1$, $X_2$ satisfy \eqref{algB} (in fact, this
requires only $X_2=\lambda\V{v}\W{w}$ for some non-zero $\lambda$
\cite{derr,mal}).  Note that for a system with two semipermeable boundaries
we are now introducing an additional parameter $z$ (which is again a
fugacity, as discussed below).  For semipermeable boundary conditions on
the left (respectively right) we take $\Wn=\W{\alpha}$ (respectively
$\Vn=\V{\beta}$, thus satisfying \eqref{algSPl} (respectively
\eqref{algSPr}).  For permeable boundary conditions on the left (right) we
take $\W{}=\W{w}$ ($\V{\,}=\V{v}$); \eqref{algPl} (respectively
\eqref{algPr}) will then be satisfied due to the choice of overall constant
in the definition \eqref{eq:defx2} of $X_2$.  This completes the
construction in the case when all the matrix elements \eqref{eq:weight}
finite which, in view of \eqref{eq:ip}, is certainly true in the region
 \be\label{eq:nicereg}
\alpha+\beta>1,\quad \alpha+v>1,\quad w+\beta>1,\quad  \hbox{and}\quad w+v > 1.
 \ee

To extend the discussion to all parameter values we must distinguish two
cases.

 \smallskip\noindent
 {\bf Case 1: One or both boundaries permeable.} In this case all weights
\eqref{eq:weight} have the form $QP(1/\alpha,1/\beta,1/w,1/v)$, with $P$ a
polynomial having positive coefficients and $Q$ a common factor depending
only on the type of system under consideration: $Q=(w+v-1)^{-1}$ in the case
P/P (that is, two permeable boundaries), $Q=(w+\beta-1)^{-1}$ in the case
SP/P, and $Q=(\alpha+v-1)^{-1}$ in the case P/SP.  The factor $Q$ cancels
in the probabilities \eqref{eq:ss} and the resulting probabilities are well
defined and positive for all positive values of the parameters; the
algebraic identities among them which express stationarity hold by analytic
continuation from the region \eqref{eq:nicereg}.

 \smallskip\noindent
 {\bf Case 2: Both boundaries semipermeable.} In this case the number $N_2$
of type~2 particles in the system is conserved by the dynamics, and the
NESS decomposes into ergodic components corresponding to the $L+1$ possible
values of $N_2$.  Within each component an argument as in Case~1 allows one
to extend the NESS to all parameter values; however, the factor $Q$
referred to there depends on $N_2$, with $Q=Q_0\equiv(\alpha+\beta-1)^{-1}$
if $N_2=0$ and $Q=Q_*\equiv(w+v-1)(\alpha+v-1)^{-1}(w+\beta-1)^{-1}$ if
$N_2>0$.  One may also consider grand canonical ensembles in which $N_2$
fluctuates.  The algebraic conditions \eqref{algB}--\eqref{algSPr} do not
determine the relative weights of configurations with different values of
$N_2$, but the particular choice of representation of the algebra discussed
above does, and thus defines a grand canonical ensemble in the region
\eqref{eq:nicereg} via \eqref{eq:ss}--\eqref{eq:Z}.  In general the
inequality of $Q_0$ and $Q_*$ is an obstruction to extending this ensemble
to all values of the parameters---negative probabilities can arise.  This
difficulty can be avoided, and a grand canonical NESS defined, by excluding
the component with $N_2=0$ from the superposition. Alternatively, a grand
canonical NESS including all components can be defined when $Q_0=Q_*$, that
is, when $w=\alpha$ or $v=\beta$.

\begin{rem}\label{rem:common} (a) The cases of semipermeable and permeable
boundary conditions are not mutually exclusive; on the left, the case $a=0$
of permeable conditions corresponds to the case $\alpha=w$ of semipermeable,
with a similar common case on the right.  

 \smallskip\noindent
 (b) The system with permeable boundaries on both ends was studied, using
other methods, in \cite{ds}.  The authors took (in our language)
$a=b=vw/(1+vwz)$ and did not observe that the NESS is in fact independent
of $a$ and $b$.

\end{rem}

As noted above, the case of two semipermeable boundaries with $w=v=1$ was
studied, in the canonical ensemble with fixed $N_2$, in \cite{arita0} and
\cite{als}, and it is easy to see that several features discussed there are
still present for arbitrary $v$ and $w$.  The
current $J_2$ is always zero.  The operator $X_2$ has rank one
(see \eqref{eq:defx2}) which implies that a type~2 particle at site $i$
decouples sites to the left of $i$ from those to the right, except that the
total number of type~2 particles in these two  subsystems must be $N_2-1$.  As a
consequence \cite{als} the marginal measure on configurations of type~2
particles is an equilibrium measure for a certain Hamiltonian with
nearest-particle interactions.   In any of the grand canonical ensembles
discussed under Case~2 above a second class particle completely decouples
the system; here we use the fact that the grand canonical ensembles for the
left and right subsystems are well defined (see the last remark under
Case~2).

Let us consider then the case in which there are permeable boundary
conditions at one or both ends of the system.  As noted above the NESS is
then independent of the parameters $a$ and/or $b$, and so is determined
by $v$, $w$, $z$, and $\alpha$ (respectively $\beta$) if semipermeable
boundary conditions are used at the left (respectively right) boundary.  In
particular, we may achieve this steady state with $a=0$ or $b=0$, so that
certainly we still have $J_2=0$. That is, all models with the rates of
Table~\ref{table:rates3} have zero current of type~2 particles.

 In all of the grand canonical ensembles discussed above the parameter $z$
plays the role of a fugacity for the type~2 particles.  To see this,
suppose that $\tau$ is a configuration with $N_2$ type~2 particles and note
from \eqref{algSPl} and \eqref{algSPr} that the weight assigned to $\tau$
by the algebra above will contain a factor $z^{N_2}$.  Now let
$w_{L,N_2}(\tau)$ denote the weight assigned to $\tau$ by the algebra
\eqref{algB}--\eqref{algSPr} of two {\it semipermeable} boundaries with
$\alpha=w$ and $\beta=v$ (for $v=w=1$ this is the weight assigned to $\tau$
in the algebra of \cite{als}).  Then we see that
 $$\mu_L(\tau)= Z_L^{-1}\sum_{N_2=0}^Lz^{N_2} w_{L,N_2}(\tau).$$
 That is, $\mu_L$ is just the ``grand canonical'' ensemble, with fugacity
$z$ for the type~2 particles, for the semipermeable system with entry
and exit rates $w$ and $v$, respectively, in the sense that this term is
often used when discussing matrix ansatz models.  It then follows, for
example, that the type~2 particles are in a grand-canonical version
of the equilibrium ensemble mentioned above.
 
\subsection{A simple special case}\label{sec:speccase}

 We close this section with the consideration of a special case in which
various quantities can be computed rather easily: we take permeable boundary
conditions on the left and semipermeable on the right, and specialize to
$v=1$ (there is a corresponding analysis with left and right reversed and
$w=1$).  The algebra  then reduces to
 \be\label{newalg}\begin{gathered}
   X_1X_0=X_1+X_0,\qquad X_1X_2=X_2,
   \qquad X_2X_0=\frac1wX_2,\\
   X_1|V\rangle=\frac1\beta|V\rangle,
     \quad \langle W|X_0=\frac1w\langle W|,\quad
\langle W|X_2=z\langle W|,\end{gathered}\ee
 and is realized by taking $\V{\,}=\V\beta$, $\W{}=\W{w}$, and
$X_2=z\V1\W{w}$ If we now define new matrices $X_1'$ and $X_2'$ by
$X'_1=X_1$ and $X'_0=X_0+X_2$ then these satisfy
 \be\label{eq:redalg}
   X'_1X'_0=X'_1+X'_0,
   \qquad 
   X'_1|V\rangle=\frac1\beta|V\rangle,
     \qquad \langle W|X'_0=\frac{1+wz}{w}\langle W|.
 \ee
 This is the algebra of an open one-species TASEP system \cite{dehp} with
entry rate $\alpha=w/(1+wz)$ and exit rate $\beta$, which implies that if
we color the type~1 particles black and the holes and type~2 particles
white, as for the $(1,2+0)$ system discussed in Section~\ref{sec:color},
then the marginal of the NESS on the black/white system is precisely the
NESS of this 1-TASEP system.  (The situation here is somewhat different
from that of Section~\ref{sec:color}, however; the correspondence here is
on the level of measures rather than dynamics, since the rate at which a
type~1 particle enters the system depends on whether there is a hole or a
type~2 particle on site 1.)

Many properties of the 2-TASEP may now be obtained.  The current and
density profile of type~1 particles are given by \eqref{currtasep}
and \eqref{denstasep}:
 \be\label{eq:simple}
  J_1=\tilde J(\alpha,\beta),\qquad \rho_1=\tilde \rho(\alpha,\beta).
 \ee
 The current of type~2 particles is zero since the right boundary
conditions are semipermeable, and the density profile of type~2 particles,
 \be\label{eq:dens2def}
\rho_2(x)=\lim_{L\to\infty,i\sim xL}\mathop{\rm Prob}(\tau_i=2),
 \ee
  may be obtained from the fact that $X_2$ is a rank one matrix:
 \begin{eqnarray}
 \mathop{\rm Prob}(\tau_i=2)
  &=&\frac{\W{}(X_0+X_1+X_2)^{i-1}X_2(X_0+X_1+X_2)^{L-i}\V{\,}}
  {\W{}(X_0+X_1+X_2)^L\V{\,}}   \nonumber \\
  &=&\label{denstwo} 
  z\frac{Z_{i-1}^{\alpha,1}Z_{L-i}^{\alpha,\beta}}{Z_L^{\alpha,\beta}},
 \end{eqnarray}
 and the known asymptotics \cite{dehp} of  $Z_L^{\gamma,\delta}
   =\W{\gamma}(X_0+X_1)^L\V{\delta}$.  
 
As an example we discuss the case $\alpha=\beta<1/2$; for convenience
we will eliminate $z$ and $\beta$ and use $\alpha$ and $w$ as basic
variables to describe the system in this case.  We will need the asymptotic
formulas \cite{dehp}
 \be\label{eq:asymp}
Z_L^{\gamma,\delta} \simeq\begin{cases}
   \ds\frac{L(1-2\gamma)^2}{\gamma^L(1-\gamma)^{L+2}},&
  \text{if $\gamma=\delta<1/2$},\\\noalign{\smallskip}
   \ds\frac{\delta(1-2\gamma)}{(\delta-\gamma)\alpha^L(1-\alpha)^{L+1}},&
   \text{if $\gamma<\delta$ and $\gamma<1/2$.}\end{cases}
 \ee
In this case all density
profiles are linear:
 \be
\rho_i=\text{linear}(r_i,r_i'), \quad i=0,1,2, 
 \ee
 where
 \begin{align}
r_1&=\alpha,\quad& r_2&=\frac{(1-2\alpha)(w-\alpha)}{w(1-\alpha)}, \quad&
    r_0&=\frac{\alpha(1-2\alpha+\alpha w)}{w(1-\alpha)}\\
r_1'&=1-\alpha,\quad& r_2'&=0,\quad&r_0'&=\alpha.
 \end{align}
 For $\rho_1$ this follows from \eqref{eq:simple}, for $\rho_2$ from
\eqref{eq:dens2def} and \eqref{denstwo}, using $z=(w-\alpha)/(w\alpha)$ and
\eqref{eq:asymp}, and then for $\rho_0$ from $\sum_i\rho_i=1$.

 We now show that these linear profiles arise, as usual, from the
occurrence of a shock in the typical (not averaged) density profiles: a
randomly located point at which (in general) all three typical profiles are
discontinuous.  In this case the shock location is marked by the position
of the rightmost type~2 particle in the system; the probability that this
particle is located at site $i$ is, from \eqref{eq:asymp},
 \be\label{eq:rmt2} 
 \frac{\W{}(X_0+X_1+X_2)^{i-1}X_2(X_0+X_1)^{L-i}\V{\,}}
  {\W{}(X_0+X_1)^L\V{\,}}
  =  z\frac{Z_{i-1}^{\alpha,1}Z_{L-i}^{w,\alpha}}{Z_L^{\alpha,\alpha}}
 \simeq\frac1L,
 \ee
 so that this particle (and hence, as we will see below, the shock) is
uniformly located throughout the system.  When conditioned on the location
$i$ of this particle the NESS decomposes into the product of the NESS's of
two independent subsystems, and by finding the densities in these
subsystems we exhibit the discontinuity in the overall (conditioned)
density profile that is the characteristic of a shock:

 \smallskip\noindent
{\bf Left subsystem:} On sites $1\ldots,i-1$ the NESS is that of a
semipermeable/permeable 2-TASEP like the full system, but with exit rate
$\beta=1$.  One may again introduce a $(1,2+0)$ coloring for this system
for which the type~1 particles form an open system with entrance rate
$\alpha$ and exit rate $1$; the colored system is in its low density phase
\cite{dehp} and thus has (up to boundary effects) a uniform density of
type~1 particles: $\rho_1\equiv\alpha=r_1$.  The density profile $\rho_2$
can then be calculated from \eqref{eq:dens2def} and \eqref{denstwo} with
the replacement $\beta\to1$; the result is a constant profile with value
$r_2$.  The profile $\rho_0$ is then also constant with $\rho_0\equiv r_0$.

 \smallskip\noindent
{\bf Right subsystem:} On sites $i+1\ldots,L$ the NESS is that of an open
one-component TASEP with entry rate $w$ and exit rate $\beta=\alpha.$ Since
$\alpha<w$ this system is in its high density phase; density profiles are
constant with $\rho_1\equiv1-\alpha$ and $\rho_0\equiv\alpha$.

Note that if $w=1$ then $r_0=r_0'$, so that the density of holes is
constant through out the system and the densities of first and second class
particles have compensating discontinuities at the shock position.
Situations of this kind were encountered in Section~\ref{sec:color} (see
Remark~\ref{rem:tables}(d)) and in the study of the ``fat shock'' in the
system with $v=w=1$ and two semipermeable boundaries \cite{als}.  
For general $w$ all three densities are discontinuous at the shock
positions, which so far as we know is a new phenomenon.  

The shock is stationary in the sense that it has no net drift velocity,
although its position does of course fluctuate.  Let us write
$J_i(\vec\rho;v,w)$ for the current of species $i$ in a uniform steady
state of an infinite system with densities
$\vec\rho=(\rho_1,\rho_2,\rho_3)$, where the exchange rates in the bulk are
1, $v$, and $w$, as in Table~\ref{table:rates1}.  The condition for a
stationary shock as described above is then that
 \be\label{eq:sshock}
J_i(\vec r;1,w)=J_i(\vec r\,';1,w) \quad\hbox{for $i=0,1,2$.}
 \ee
 It would be useful to have analytic expressions for the currents
$J_i(\vec\rho;v,w)$ so that these relations could be checked; such expressions
were obtained for the case $v=w=1$ in \cite{djls}, essentially by a coloring
argument.  However, for any given $\vec\rho$, $v$, and $w$ one may compute
$J_i(\vec\rho;v,w)$ with good accuracy by Monte Carlo simulation of the uniform
system on a large ring; we have done so and thus verified that
\eqref{eq:sshock} holds for a variety of choices of $\alpha$ and $w$.

\begin{rem}\label{rem:mf} A simple mean field calculation of the currents
in a uniform two species system yields
$J_1(\vec\rho;v,w)=\rho_1(\rho_0+v\rho_2)$,
$J_2(\vec\rho;v,w)=\rho_2(w\rho_0-v\rho_1)$, and
$J_0(\vec\rho;v,w)=\rho_0(-\rho_1-w\rho_2)$.  These mean field values are in
fact correct when $v=w=1$, as is well known, but the Monte Carlo
simulations referred to above show that they are not correct in general.
\end{rem}

\end{document}

\bibitem{derr2} B. Derrida,An exactly soluble non-equilibrium system: The
asymmetric simple exclusion process, {\it Physics Reports} {\em 301}
(1998), 65--83.
 
\bibitem{spitz} F.~Spitzer, Interaction of Markov processes,
    {\it Advances in Math.} {\bf 5} (1970), 246--290.

\bibitem{ligg1} T.~M.~Liggett, {\em Interacting particle systems},
Springer-Verlag, New York, 1985.

\bibitem{ddm} B.~Derrida, E.~Domany, and D.~Mukamel, An exact solution of
a one-dimensional asymmetric exclusion model with open boundaries, {\em J.
Stat. Phys.} {\bf 69} (1992), 667--687.

\bibitem{sd} G. ~Sch\"utz and E.~Domany, Phase transitions in an exactly
soluble one-dimensional asymmetric exclusion model, {\em J. Stat. Phys.}
{\bf 72} (1993), 277-296.

\bibitem{sch} G.~M.~Sch\"utz, Exactly solvable models for many-body systems
  far from equilibrium,  in {\it Phase Transitions and Critical Phenomena,
    Vol 19}, ed. C.~Domb and J.~L.~Lebowitz, Academic Press, London, 2000.

\bibitem{kjs} K.~Krebs, F.~H.~Jafarpour and G.~M.~Sch\"utz, Microscopic
structure of travelling wave solutions in a class of stochastic interacting
particle systems, {\em New J. Phys.} {\bf 5} (2003), 145.1-145.14.

\bibitem{dls1} B.~Derrida, J.~L.~Lebowitz, and E.~R.~Speer, Shock profiles
for the asymmetric simple exclusion process in one dimension, {\em J. Stat.
Phys.} {\bf 89} (1997), 135--167.

\bibitem{feller} W.~Feller, {\em An Introduction to Probability Theory and
  its Applications II, $2^{\it nd}$ edition},  John Wiley \& Sons, New
  York, 1971.

\bibitem{cattri} N.~J.~A.~Sloane, {\em The On-Line Encyclopedia of
Integer Sequences}, {\bf A009766},
%\href{http://www.research.att.com/~njas/sequences/A009766}
{\texttt{http://www.research.att.com/$\sim$njas/sequences/A009766}}.

\bibitem{hill} T.~L.~Hill, {\em Statistical Mechanics: Principles and
Selected Applications}, McGraw-Hill, New York, 1956.

\bibitem{percus} J.~K.~Percus, Exactly solvable models of classical
many-body systems,  in {\it Simple models of equilibrium and nonequilibrium
  phenomena}, ed. J.~L.~Lebowitz, Amsterdam, North-Holland, 1987.

\bibitem{kuh} M.~Kac, G.~E.~Uhlenbeck, and P.~C.~Hemmer, On the van der
  Waals theory of the vapor-liquid equilibrium. I. Discussion of a
  one-dimensional model, {\em J.~Math. Phys.} {\bf 4} (1963), 216--228.

\bibitem{zrp} M.~R.~Evans and T.~Hanney, Nonequilibrium statistical
mechanics of the zero-range process and related models. {\it J.~Phys. A:
  Math Gen.} {\bf 38} (2005), R195--R240.

\bibitem{bdgjl} L.~Bertini, A.~De Sole, D.~Gabrielli,
G.~Jona-Lasinio, and C.~Landim, Stochastic interacting particle systems out
of equilibrium,  {\em J. Stat. Mech.} (2007), P07014, 35pp.

\bibitem{fm} P.~A.~Ferrari and J.~B.~Martin, Stationary Distributions
  of Multi-Type Totally Asymmetric Exclusion Processes,  {\it Ann. Prob.}
  {\bf 35} (2007), 807--832.

\bibitem{kls} S.~Katz, J.~L.~Lebowitz, and H.~Spohn, Non-equilibrium
Steady States of Stochastic Lattice Gas Models of Fast Ionic Conductors,
{\em J.  Stat. Phys.} {\bf 34} (1984), 497--537.

\bibitem{dls} B.~Derrida, J.~L.~Lebowitz, and E.~R.~Speer, Exact large
deviation functional of a stationary open driven diffusive system: the
asymmetric exclusion process, {\em J. Stat. Phys.} {\bf 110} (2003), 
775--810.

\bibitem{de} B.~Derrida and  M.~R.~Evans, Bethe ansatz solution for a
  defect particle in the asymmetric exclusion model,
{\em J. Phys. A} {\bf 32} (1999), 4833--4850.

\end{thebibliography}

\end{document}